# On the quest for the strongest materials

Javier LLorca[1,2]

[1]IMDEA Materials Institute, C/Eric Kandel 2, 28906 – Getafe, Madrid, Spain.
[2]Department of Materials Science, Polytechnic University of Madrid. E. T. S. de Ingenieros de Caminos, 28040 – Madrid, Spain.

Looking for the strongest material has always been a scientific goal for the structural materials community. Early theoretical developments[1] showed that the maximum attainable strength in tension (shear) was controlled by the fracture of the interatomic bonds and was of the order of ≈ $E/10$ (≈ $G/10$), where $E$ and $G$ stand for the elastic and shear moduli, respectively. It was also evident that it was not trivial to achieve these strengths because the defects in the solid will lead to the inelastic relaxation of the lattice strains or to the brittle fracture well before the atomic bond could be stretched up to the theoretical limit. As a result, the maximum elastic tensile strains supported by bulk solids are of the order 0.2%-0.4% while tensile strains of up to 3-4% were measured in μm-size whiskers[2]. Nevertheless, recent progress in the synthesis of nanoscale objects as well as in nanomechanical testing techniques opened the possibility to probe the strength of material systems which were practically free of defects and opened the quest for the strongest material[3]. In parallel, atomistic simulations based on the density-functional theory and molecular dynamics have shown their ability to predict accurately the fracture strength of perfect crystals as well as to explore the influence of defects and free surfaces on this magnitude[4].

Because the C-C covalent bond is the strongest one in nature, the search for the strongest material has been focused in 1D carbon nanotubes (CNTs) and 2D graphene nanoscale objects and a summary of recent results can be found in Fig. 1. Experimental resuls and *ab initio* calculations indicate that the elastic modulus of both CNTs and graphene is ≈ 1 TPa[5-8] and, thus, the data in Fig. 1 show strength values very close to the theoretical limit for 1D and 2D carbon materials with $sp^2$ covalent bonds. However, the reported tensile strength of bulk cubic diamond with $sp^3$ covalent bonds is much smaller (<10 GPa), although the stiffness is equivalent to that of $sp^2$ carbons[9]. These differences in strength were partially attributed to the difficulties associated with the development of brittle fracture from defects during tensile deformation of bulk samples. Higher tensile strengths (up to 20 GPa) for diamond have been reported from Hertzian indentation tests[10] but these values have to be analyzed with caution because of the uncertainties associated with the spherical indentation to determine the tensile strength. Regardless of the experimental technique, fracture of diamond occurs by cleavage along the (111) plane which presents the lowest fracture energy.

Banerjee *et al.*[11] reports in this issue of Science experimental results of the elastic deformation of nanoscale single-crystal diamond needles which failed when the maximum local tensile stress was around ≈ 89-98 GPa. This magnitude (which corresponds to a tensile strain of ≈ 9%) is very close to the theoretical limit for diamond and to the maximum strengths values reported for CNTs and graphene. These large elastic strains were applied through bending of the nano-needles and fracture was not triggered at smaller strains due to the small volume of the nanoscale needles, the paucity of defects and the smooth surface, which was produced by careful reactive ion etching from a <111> oriented CVD diamond film. Fracture occurred by cleavage along (111) planes and the strength measured was supported by combined density functional theory/molecular dynamics simulations at 300K, which predicted a maximum critical strain of 13% associated to a tensile strength of 130 GPa.

The results in Banerjee et al.[11] are relevant from the fundamental viewpoint because they show that the strength of bulk diamond with $sp^3$ bonds is equivalent to that reported for CNTs and graphene with $sp^2$ bonds. But, in addition, they open the possibility of modifying fundamental properties of diamond by means of elastic strain enginnering[12]. The properties of any crystalline material depend on the lattice parameters, which are dictated by the electronic structure. Large elastic deformations (around 10%) modify the electronic structure, leading to changes of electronic, magnetic, catalytic, etc. properties that could be even tuned as a function of the applied strain tensor. This concept is not new and has been successfully applied to enhance the drive current in complementary metal-oxide semiconductors technology by improving the electron and hole mobility through lattice straining[13]. Moreover, many other applications have been reported[12], including for instance, the transformation of paramagnetic materials into ferromagnetic ones[14,15], or the manipulation of the mechano-chemistry coupling to increase the catalytic activity[16]. Nevertheless, the elastic strains applied to modify the properties in all these cases were much lower (< 4-5%) than those reported above for CNTs, graphene and diamond needles, which were closer to 10%. Changes in properties should increase with the elastic strain and, as postulated by Gilman[17], the electronic structure should be drastically modified near the critical bond breaking strain, leading to unusual or singular chemical and physical properties. Thus, exploration of the effect of very large elastic strains (up to 10%) on the properties of crystalline solids may lead to the discovery of new or unexpected behaviors.

It should be finally noted that many applications of elastic strain engineering require that the elastic strains are distributed over of significant area or volume. This can achieved by means of epitaxial growth[13,14,16] or severe plastic deformation[15] in the case of bulk materials. However, this task is more difficult in the case of nanoscale objects, such as CNTs and graphene sheets, because of the problems to maintain the mechanical continuity between different objects. For instance, the maximum strengths reported for CNT fibers (made up by bundles of CNTs) are of the order of 2.5 GPa, well below the values reported for single CNTs[18-19] (Fig. 1). Elastic straining techniques that can be applied to bulk materials, such as the elastic bending of nanoneedles in Banerjee et al.[11] can be potentially extended to large surfaces and different materials and can be used to explore the effect of large elastic strains on chemical and physical properties of diamond and other solids.

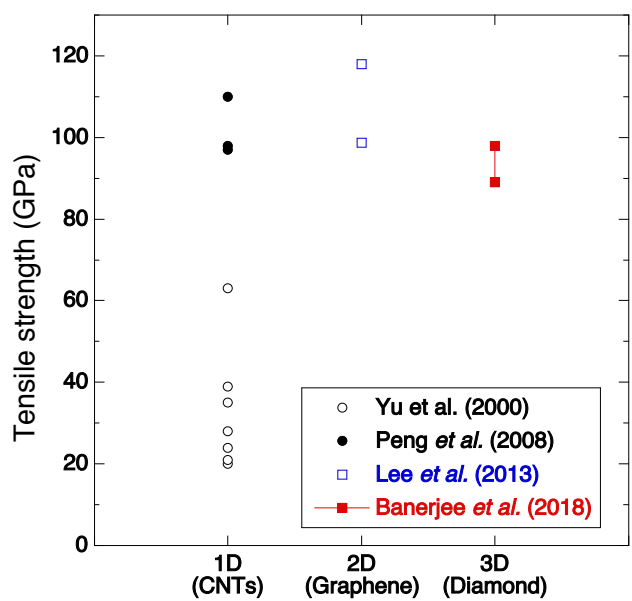

Fig. 1. Experimental results of the tensile strength of multiwall carbon nanotubes (CNTs)[5,6], graphene[8] and diamond nanoneedles[11].